\documentclass[english,12pt]{article}

\usepackage[LGR,T1]{fontenc}
\usepackage[utf8]{inputenc}
\usepackage{geometry}
\usepackage{array}
\usepackage{alltt}
\usepackage{float}
\usepackage{booktabs}
\usepackage{textcomp}
\usepackage{multirow}
\usepackage{amsmath}
\usepackage{amsthm}
\usepackage{amssymb}
\usepackage{authblk}
\usepackage{sectsty}
\usepackage{hyperref}
\usepackage{graphicx}
\usepackage{blindtext}
\usepackage{caption}
\usepackage{subcaption}
\usepackage{lmodern}
\usepackage{multicol}
\usepackage{tabularray}

\newcommand{\Ebb}{\mathbb{E}}
\newcommand{\Ibb}{\mathbb{I}}
\newcommand{\Pbb}{\mathbb{P}}
\newcommand{\Vbb}{\mathbb{V}}

\usepackage{tikz}
\usetikzlibrary{arrows,automata}
\tikzset{
	semi/.style={
		semicircle,
		draw,
		minimum size=2em
	}
}

\usetikzlibrary{shapes, snakes, graphs, shapes.geometric, positioning}

\sectionfont{\fontsize{14}{13}\selectfont}
\subsectionfont{\fontsize{12}{12}\selectfont}

\allowdisplaybreaks

\makeatletter

\ProvideTextCommand{\~}{LGR}[1]{\char126#1}

\theoremstyle{plain}

\usepackage{amsmath}

\makeatother

\usepackage{babel}
\providecommand{\theoremname}{Theorem}

\usepackage[
backend=biber,
style=nature,
uniquename=init,
giveninits
]{biblatex}

\begin{document}

\title{Simplifying Causal Mediation Analysis for Time-to-Event Outcomes using Pseudo-Values}











\author[1$\dag$]{Alex Ocampo}
\author[1]{Enrico Giudice}
\author[1]{Dieter A. H\"aring}
\author[2]{Baldur Magnusson}
\author[3]{Theis Lange}
\author[4,5]{Zachary R. McCaw}

\affil[1]{Novartis Pharma AG, Basel, Switzerland}
\affil[2]{UCB Farchim SA, Bulle, Switzerland}
\affil[3]{Department of Biostatistics, University of Copenhagen}
\affil[4]{insitro, South San Francisco, USA}
\affil[5]{Department of Biostatistics, University of North Carolina at Chapel Hill, Chapel Hill, NC, USA }
\affil[$\dag$]{Email: alex.ocampo@novartis.com}
\setcounter{Maxaffil}{0}
\renewcommand\Affilfont{\itshape\small}

\date{}
\maketitle
\vspace{-11mm}
\begin{abstract}
Mediation analysis for survival outcomes is challenging. Most existing methods quantify the treatment effect using the hazard ratio (HR) and attempt to decompose the HR into the direct effect of treatment plus an indirect, or mediated, effect. However, the HR is not expressible as an expectation, which complicates this decomposition, both in terms of estimation and interpretation. Here, we present an alternative approach which leverages pseudo-values to simplify estimation and inference. Pseudo-values take censoring into account during their construction, and once derived, can be modeled in the same way as any continuous outcome. Thus, pseudo-values enable mediation analysis for a survival outcome to fit seamlessly into standard mediation software (e.g. \texttt{CMAverse} in \texttt{R}). Pseudo-values are easy to calculate via a leave-one-observation-out procedure (i.e. jackknifing) and the calculation can be accelerated when the influence function of the estimator is known. Mediation analysis for causal effects defined by survival probabilities, restricted mean survival time, and cumulative incidence functions - in the presence of competing risks - can all be performed within this framework. Extensive simulation studies demonstrate that the method is unbiased across 324 scenarios/estimands and controls the type-I error at the nominal level under the null of no mediation. We illustrate the approach using data from the PARADIGMS clinical trial for the treatment of pediatric multiple sclerosis using fingolimod. In particular, we evaluate whether an imaging biomarker lies on the causal path between treatment and time-to-relapse, which aids in justifying this biomarker as a surrogate outcome. Our approach greatly simplifies mediation analysis for survival data and provides a decomposition of the total effect that is both intuitive and interpretable. \\
\; \\
\textit{Keywords}: Mediation, Pseudo-Values, Time-to-event, Survival Analysis, Restricted Mean Survival Time, Competing Risks
\end{abstract}


\section{Introduction}


Survival outcomes are relevant in many practical applications but they pose difficulties for mediation analyses both in terms of defining the problem as well as for performing estimation and inference. In this article, we aim to explore and overcome these challenges by leveraging pseudo-values to provide an intuitive and interpretable alternative to existing mediation approaches in survival contexts. The pseudo-values can be easily applied in a mediation analysis for time-to-event endpoints and implemented for summary measures that are easily interpretable such as the survival probability, restricted mean survival time, and cumulative incidence in the presence of competing risks. We thereby fill a gap in the existing mediation literature for survival outcomes by offering a simple yet theoretically sound approach.

Most mediation analyses with survival outcomes focus on decomposing hazards ratios \cite{rochon2014mediation,lee2014depressive,fritz2015mediation}. When first approaching mediation in a survival context, a natural starting point is the regression based difference method, whereby two Cox-proportional hazards models are fit with and without the candidate mediator as a covariate and the difference between the exposure coefficients is measured. There are a number of drawbacks to this approach. First, performing mediation analysis with the Cox model is mathematically inconsistent because proportional hazards cannot simultaneously hold for the Cox models with and without the mediator \cite{lange2011}. Secondly, the mediation analysis does not have a causal interpretation, mainly due to the non-collapsibility of the hazard ratio, which changes the estimand targeted by the exposure coefficient even in the absence of mediation \cite{daniel2021making}. This, in effect, invalidates any mediation based on two Cox models. 

More refined approaches avoid the above issues by only requiring estimation of a single Cox model. One strategy is based on weighting. For example, the inverse odds ratio weighting (IORW) approach \cite{tchetgen2013inverse} fits a logistic model for a binary exposure regressed on the mediator, then uses the logistic model to build weights that are incorporated into a single Cox model for the survival outcome. Contrasting this weighted model with the total effect model enables estimation of the mediation effects. This removes the need for two Cox models that are fundamentally at odds with one-another. Besides the IORW approach, one can use a marginal structural model approach to apply mediation weights to a Cox model that decomposes the HR \cite{lange2012simple}. Yet another strategy employs imputations of the potential outcomes used to define the natural direct and indirect effects. Imputations from a parametric model (e.g.\@ Weibull) are supplied to a single Cox model, with which the mediated and non-mediated effects are estimated \cite{lange2017applied}. This last approach is similar to the imputation approach of the \texttt{medflex} package in \texttt{R}, although the latter cannot handle survival outcomes \cite{steen2017medflex}. Still, these methods are quite technical in their implementations for the applied analyst and suffer from the usual limitations of Cox models, such as requiring proportional hazards \cite{uno2014rmst, mccaw2021interpretable}. 

Another way to avoid the problems of mediation analysis with Cox models is to instead adopt either a semi-parametric additive hazards model or parametric accelerated failure time model. Since both these models are linear in nature, casting them into a mediation framework is straightforward. Thus, many papers have proposed their use in mediation analysis for survival outcomes \cite{aalen2020time, lange2011, russell2020mediation, kormaksson2024dynamic}. There exists both theoretical and simulated evidence that using accelerated failure time models for mediation has advantages over Cox models \cite{gelfand2016mediation}. The additive hazards model also avoids making a proportional hazards assumption \cite{lin1997additive}. 
These advantages - linearity and the absence of a proportional hazards requirement - are shared by our proposed approach as well, but we further build on these desirable features by allowing the analyst to quantify effects in terms of probabilities and time, which are translatable in layperson's terms.

To achieve our objective in simplifying mediation analysis with time-to-event outcomes, we will leverage pseudo-values. As first studied by Andersen in \cite{andersen2003}, pseudo-values enable regression analysis of any survival functional to be conducted using familiar linear models, which are intuitive and easy to estimate via standard software. Later, Andersen and colleagues extended the use of pseudo-values to causal inference \cite{andersen2017causal}. Regressions for time-to-event outcomes based on pseudo-values have advantageous inferential properties, most notably that it exhibits double robustness, which many standard methods in the time-to-event outcome setting do not \cite{wang2018simple, gabriel2024propensity}. The first proposal, to our knowledge, of using pseudo-values for mediation was in the Ph.D.\@ thesis of Chernofsky in 2022 \cite{chernofsky2022methods} where the approach was applied to decompose an effect on restricted mean survival time (RMST). A similar application of pseudo-values has been developed for instrumental variable analyses \cite{kjaersgaard2016instrumental}. This paper goes beyond existing work by developing pseudo-value-based mediation analysis for new endpoints, including the survival and cumulative incidence probabilities, and indeed the approach is applicable to any estimand that can be expressed as an expectation. 



We apply our methodology to conduct a mediation analysis using the data of the randomized, phase III, PARADIGMS clinical trial, which evaluated the efficacy and safety of fingolimod versus interferon $\beta$-1a for the treatment of pediatric multiple sclerosis (MS). An important outcome of the study was to assess the time-to-first clinical relapse (defined by the appearance of new or worsening neurological symptoms due to MS related inflammation in the brain).  As a biomarker, we consider the number of gd-enhanced T1 brain lesions, a measure of focal inflammation of the brain, as assessed by  magnetic resonance imaging (MRI). Establishing these MRI lesions as a mediator of fingolimod's efficacy on relapses in MS would not only clarify the treatment's mechanism of action, but also support this biomarker's use as a surrogate endpoint in future studies \cite{joffe2009related}. 

The remainder of this paper is organized as follows: section 2 introduces our notation and gives a review of the necessary background knowledge on mediation, survival estimators, pseudo-values, and influence functions, all of which are employed in our approach. Section 3 provides step-by-step instructions to implement pseudo-value-based mediation analysis for time-to-event outcomes. Section 4 presents simulation studies demonstrating that the approach is unbiased for all causal effects of interest and provides valid inference, with proper control of the type I error and well-calibrated coverage probabilities. Section 5 applies the method to the PARADIGMS clinical trial for the treatment of MS. We conclude in section 6 with a discussion.  

\section{Notation and Background}
\subsection{Mediation}

\begin{figure}[H]
\centering
{\includegraphics[width=0.9\columnwidth]{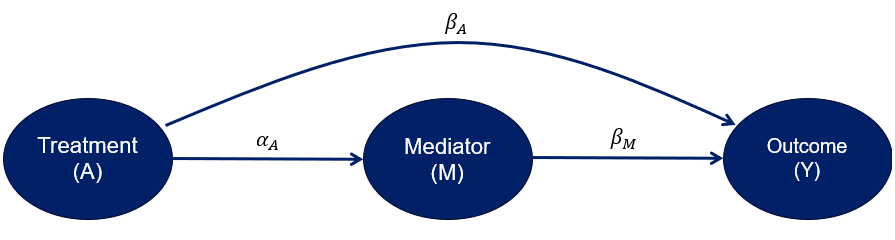}}
\caption{Causal Graph for Mediation. $\beta_{A}$ characterizes the direct effect of the treatment (A) on the outcome (Y). $\alpha_{A}$ constitutes the effect of the treatment on the mediator, and $\beta_{M}$ the effect of the mediator on the outcome. When outcomes are continuous and modeled linearly, the product $\alpha_{A}\beta_{M}$ constitutes the mediated effect, and the sum of the direct and mediated effects $\beta_{A} + \alpha_{A}\beta_{M}$ comprises the total effect the treatment on the outcome.}
\label{fig:causal-mediation-graph}
\end{figure}

Mediation analysis helps explain causal mechanisms by quantifying pathways through which a causal effect flows. This can be useful in characterizing the biological mechanisms of action responsible for a given treatment's clinical benefit. For continuous endpoints, mediation analysis by means of linear models has a straightforward and well-known decomposition, popularized in the 1986 paper by Baron $\&$ Kenny \cite{baron1986moderator}. To see how, one can depict the mediation setup with a casual graph. We present such a graph considering a binary randomized treatment ($A$, also sometimes referred to as exposure), a mediator ($M$) and an outcome ($Y$) in Figure \ref{fig:causal-mediation-graph}. 

In the above graph, the pathways relevant to a mediation analysis are labeled by Greek letters. When all variables are continuous, these effects can be estimated using the following linear models:
\begin{align}
    \text{Outcome Model}:\quad \Ebb[Y|Z,M] &= \beta_0 + \beta_AA + \beta_MM, \label{eqn:outcome-model} \\
    \text{Mediation Model}:\quad \Ebb[M|A] &= \alpha_0 + \alpha_AA. \label{eqn:mediator-model}
\end{align}

The parameters of these models correspond to the paths in Figure \ref{fig:causal-mediation-graph}. Using potential outcome notation, let $M(a)$ denote the value of the mediator $M$ when the treatment is set to $A=a$, and let $Y(a, m)$ denote the value of the outcome $Y$ when the treatment is set to $A=a$ while the mediator is set to $M=m$. Under the assumptions of linearity and no interaction, models (\ref{eqn:outcome-model}) and (\ref{eqn:mediator-model}) can be used to estimate the following causal effects:
\begin{align*}
    \text{Natural direct effect}:\quad \text{NDE} &= \Ebb[Y\{1,M(0)\}]-\Ebb[Y\{0,M(0)\}] = \beta_A, \\
    \text{Natural indirect effect}:\quad \text{NIE} &= \Ebb[Y\{1,M(1)\}]-\Ebb[Y\{1,M(0)\}] = \alpha_A\beta_M, \\
    \text{Total effect}:\quad \text{TE} &=\Ebb[Y\{1,M(1)\}]-\Ebb[Y\{0,M(0)\}] = \beta_A + (\alpha_A\beta_M).
\end{align*}

 This decomposition, due to Baron $\&$ Kenny \cite{baron1986moderator}, formalized mediation and provided an intuitive graphical interpretation of the problem. Post Baron $\&$ Kenny, it became evident that confounders need to be considered in order to identify these effects \cite{judd1981process,vanderweele2016mediation}. Specifically, confounders of the treatment-mediator, treatment-outcome, and mediator-outcome relationships need to be included in the above models to properly estimate the mediated pathway. Also, treatment cannot affect any mediator-outcome confounder. In the setting of randomized clinical trials, the treatment is assigned at random, which removes confounding of the treatment-outcome and treatment-mediator relationships. 

 

When outcomes are not continuous, as with time-to-event endpoints, the formulas above do not hold in general. This motivated our adoption of the pseudo-value approach, which translates survival analysis to the linear model setting \cite{andersen2010}, allowing mediation analysis on survival endpoints to leverage the simple formulas above. Before introducing the pseudo-values, we will first define the survival estimands that we consider for mediation analysis with time-to-event outcomes. 


\subsection{Survival Estimands $\&$ Estimators}

This paper will consider three survival estimands: 1) the survival probability, 2) the restricted mean survival time, and 3) the cumulative incidence curve in the presence of competing risks. All of these estimands share a number of advantages. They all target interpretable and model-free quantities that can be easily communicated to non-statisticians \cite{mccaw2021interpretable}, their causal interpretations in a randomized trial are uncontroversial, and all have non-parametric estimators with desirable statistical properties \cite{andersen1993}. These estimands have also been considered for instrumental variable analysis with pseudo-values \cite{kjaersgaard2016instrumental}.

To define these estimands, let $T$ denote a subject's time to event and $C$ the time to censoring. For subject $i$, define the observation time $U_i$ and the minimum of their event and censoring times, $U_{i}=\min(T_i,C_i)$, and let $\delta_i=\Ibb(T_i\leq C_i)$ denote the event status indicator. 

The survival probability is defined as:
\begin{align*}
     S(t) = \Pbb(T > t),
\end{align*}

which is the probability of surviving past time $t$. If the event is not death, $S(t)$ is the probability of being event-free at time $t$. In the two sample-setting, let $S_{1}(\tau)$ denote the survival probability at time $\tau$ in the treatment arm, and $S_{0}(\tau)$ the survival probability in the reference arm. The effect of treatment on the survival probability at time $\tau$ can be characterized by the difference, $S_{1}(\tau) - S_{0}(\tau)$. Within each arm, the survival probability can be estimated non-parametrically via the Kaplan-Meier estimator:
\begin{align}
    \hat{S}(\tau)=\prod_{U_i\leq \tau} \left\{ 1 - \frac{dN(U_i)}{Y(U_i)} \right\},
    \label{eq:KM}
\end{align}
where $N(t)$ is the sample-level counting process representing the count of individuals who have experienced the event up to and including time $t$, and $Y(t)$ indicates the number of individuals at risk just before time $t$. The increment $dN(t)$ represents the number of new events added to the cumulative count $N(t)$ at time $t$. Although the survival probability is perhaps the most intuitive and interpretable time-to-event estimand, it is a local measure of the survival curve that can be sensitive to the choice of $\tau$. A more global measure is the restricted mean survival time \cite{uno2014rmst}.

The Restricted Mean Survival Time (RMST) is defined as:
\begin{align*}
    \mu(\tau) = \Ebb\big\{\min(T, \tau)\big\} = \int_0^\tau S(u)du,
\end{align*}

which has a geometric interpretation as the area under the survival curve up to time $\tau$. In the two-sample setting, the total effect of treatment can be quantified by the difference $\mu_1(\tau)-\mu_0(\tau)$. When estimated by means of a randomized controlled trial, this difference can be interpreted as the increase in the expected time-to event, up to time $\tau$, attributable to the experimental therapy. This time-scale summary of treatment efficacy is more interpretable for clinicians and patients than is the HR \cite{weir2019, mccaw2019}. We can estimate the RMST by integrating the Kaplan-Meier curve from (\ref{eq:KM}) from time $0$ to time $\tau$:
\begin{align}
    \hat{\mu}(\tau) = \int_0^\tau \hat{S}(u) du.
\end{align}

Our final estimand considers the presence of competing risks, which are terminal events that can occur before and preclude observation of the event of interest \cite{austin2016cr, mccaw2022cr}. In such cases, the cumulative incidence curve for a type-$j$ terminal event $j\in {1,\dots,J}$ is defined as:
\begin{align*}
    F_{j}(t) = \Pbb(T \leq t, \delta = j).
\end{align*}
As for the previous estimands, in the two-sample setting, the treatment effect can be quantified by the difference in the cumulative incidence of the event of interest $j$ at time $\tau$, $F_{j,1}(\tau) - F_{j,0}(\tau)$. Within each arm, the cumulative incidence can be estimated non-parametrically by the Aalen-Johansen estimator \cite{andersen1993}:
\begin{align}
    \widehat{F}_j(\tau) = \int_0^\tau \prod_{U_i\leq u} \left\{ 1 - \frac{\sum_{j'=1}^{J} dN_{j'}(U_i)}{Y(U_i)} \right\} \frac{dN_{j}(u)}{Y(u)}.
\end{align}
When $J = 1$, the cumulative incidence $F_{1}(t)$ reduces to $1 - S(t)$, however this relation no longer holds in the presence of $J \geq 2$ terminal events. 

So far, we have discussed estimation of survival endpoints without reference to covariates. However, adjusting for covariates is essential to performing mediation analysis. In the next section, we discuss how pseudo-values enable us to specify regression models for each of the above estimands. 

\subsection{Pseudo-values}

Pseudo-values \cite{andersen2003, andersen2010, andersen2017causal, ambrogi2022} provide a mechanism for estimating a parameter of the form:
\begin{align*}
    \theta = \Ebb\big\{g(T)\big\},
\end{align*}
where $g$ is a known function and the random variable $T$, representing the time-to-event, is potentially subject to censoring. All the survival estimands considered in the previous section can be written in this form. For instance, the survival probability is expressible as an expectation:
\begin{align*}
    S(t) = \Pbb(T > t) = \Ebb\big\{\Ibb(T > t)\big\},
\end{align*}
as is the restricted mean survival time (RMST) up to time $\tau$:
\begin{align*}
    \mu(\tau) = \Ebb\big\{\!\min(T, \tau)\big\},
\end{align*}
and the cumulative incidence of a type-$j$ terminal event:
\begin{align*}
    F_{j}(t) = \Pbb(T \leq t, \delta = j) = \Ebb\big\{\mathbb{I}(T \leq t, \delta = j)\big\}.
\end{align*}
For a given estimand $\theta$, the pseudo-value for the $i$th patient is calculated as: 
\begin{align}
    Y_{i} = n\hat{\theta} - (n-1)\hat{\theta}_{(-i)},
    \label{eqn:pseudo-value-def}
\end{align}
where $n$ is the sample size, $\hat{\theta}$ is an estimate of the target parameter based on the full sample, and $\hat{\theta}_{(-i)}$ is the estimate based on all patients except patient $i$ (i.e. the leave-one-out or jackknife estimate). When constructing the pseudo-values, established methods that account for censoring, such as the Kaplan-Meier and Aalen-Johansen estimators, are used to calculate $\hat{\theta}$ and $\hat{\theta}_{(-i)}$ \cite{andersen1993}. Once constructed, however, the pseudo-values can be modeled using familiar techniques for continuous outcomes such as linear regression or generalized estimating equations \cite{liang1986}. Moreover, the coefficients from these models target the effects of covariates on the original estimand, with the contrast estimated depending on the link function. For example, with a standard linear model employing the identity link, the coefficient on the exposure estimates the change in $\theta$ per unit change in the exposure, holding the remaining covariates constant. As the pseudo-values will represent the response variable in the outcome model (\ref{eqn:outcome-model}), we will denote them by $Y_{i}$. 

For scenarios with large sample sizes, many time-points of interest $\tau\in \mathcal{T}$, or computationally intensive estimators $\hat{\theta}$, calculating the pseudo-values via the leave-one-out estimator can become computationally intensive. We therefore describe how the influence function can be used to approximate pseudo-values.

\subsection{Influence functions}
\label{sec:influence-functions}

Suppose $X_{1}, \dots, X_{n}$ are independent and identically distributed observations with cumulative distribution function (CDF) $F$. As above, $\theta$ denotes the parameter of interest (the estimand). Let $\theta = T(F)$ express the estimand statistical functional of the CDF. This functional is then applied to data to obtain $\hat{\theta}=T(X_{1}, \dots, X_{n}) = T(\mathbb{F}_{n})$, where $\mathbb{F}_{n}$ is the empirical CDF (ECDF). The influence function $\varphi(X_i)$ of an estimator is defined as the instantaneous relative influence of $X_i$ on $T(F)$ as $\epsilon \to 0$:
\begin{equation*}
\begin{split}
    \varphi_{T,F}(X_i) & = \lim_{\epsilon\to0}\frac{T((1-\epsilon)F+\epsilon\delta_{X_i})-T(F)}{\epsilon} 
\end{split}
\end{equation*}
where $\delta_{x} = \Ibb(X<x)$ corresponds with a point-mass distribution on $x$. This formal definition considers what would happen if the distribution $F$ were $\epsilon$-contaminated by an observation at $X_i$. Formally, the influence function is the G$\hat{\text{a}}$teaux derivative of the statistical functional $T(\cdot)$ in the direction of $X_i$ \cite{serfling1980}, and can therefore be equivalently expressed using Leibniz notation as: 
\begin{equation}
\begin{split}
    \varphi_{T,F}(X_i) & = \frac{\partial}{\partial\epsilon} T((1-\epsilon)F+\epsilon\delta_{X_i})\bigg|_{\epsilon=0}
\end{split}
    \label{eqn:IF-def}
\end{equation}

The regular and asymptotically linear estimator $\hat{\theta} = T(\mathbb{F}_{n})$ admits a unique influence function expansion \cite{tsiatis2006} if there exists a function $\varphi$ with expectation zero such that:
\begin{align}
    \sqrt{n}\big(\hat{\theta} - \theta\big) = \frac{1}{\sqrt{n}}\sum_{i=1}^{n}\varphi(X_{i}) + o_{p}(1). 
\end{align}
The influence function $\varphi$ is often used as a means of estimating the asymptotic variance of an estimator:
\begin{align*}
    n \cdot \hat{\Vbb}(\hat{\theta}) = \frac{1}{n}\sum_{i=1}^{n}\varphi^{2}(X_{i}). 
\end{align*}
In addition, the influence function is also closely connected to the pseudo-value. Following \cite{boos2013}, the pseudo-value from (\ref{eqn:pseudo-value-def}) is expressible as:
\begin{align}
    Y_{i} - \hat{\theta} = -(n-1)(\hat{\theta}_{(-i)} - \hat{\theta}) = \frac{T\{(1-\epsilon_{n})\mathbb{F}_{n} + \epsilon_{n}\delta_{i}\} - T(\mathbb{F}_{n})}{\epsilon_{n}}.
    \label{eqn:pseudo-to-inf}
\end{align}
Here $\hat{\theta} = T(\mathbb{F}_{n})$ is the full sample estimate, $\hat{\theta}_{(-i)}$ is the leave-one-out estimator, $\delta_{i}$ defined above corresponds with a point-mass on observation $i$, and $\epsilon_{n} = -(n-1)^{-1}$. In the limit as $n \to \infty$, the right hand side converges to the influence function $\varphi(X_{i})$ for observation $i$ \cite{boos2013} (p.388 - 390). 
Therefore, if an expression for the influence function is known, the pseudo-value can be approximated as:
\begin{align}
    Y_{i} \approx \hat{\theta} + \varphi(X_{i}).
    \label{eqn:pseudo-approx}
\end{align}
To give the influence functions of common survival estimators, let $\{(U_{i}, \delta_{i})\}_{i=1}^{n}$ denote the observation time and status indicator for $n$ independent subjects, where $U_{i} = \min(T_{i}, C_{i})$ is the minimum of the event time $T_{i}$ and censoring time $C_{i}$, and $\delta_{i} = \Ibb(T_{i} \leq C_{i})$. Define the subject-level counting process $N_{i}(t) = \Ibb(U_{i} \leq t, \delta_{i})$ and at-risk process $Y_{i}(t) = \Ibb(U_{i} \geq u)$. The subject-level martingale is:
\begin{align*}
    M_{i}(t) = N_{i}(t) - \int_{0}^{t}Y_{i}(u)dA(u),
\end{align*}
where $A(u)$ is the cumulative hazard for the event time $T$ based on the entire sample \cite{andersen1993}. For the survival probability, the influence function is:
\begin{align*}
    \varphi_{i}(t) = -S(t)\int_{0}^{t}\frac{dM_{i}(u)}{y(u)},
\end{align*}
where $y(u) = \Pbb(U > u)$ and $S(t)$ is the survival function for $T$. For the RMST, the influence function is:
\begin{align*}
    \varphi_{i}(t) = -\int_{0}^{t}\frac{\nu_{\tau}(u)dM_{i}(u)}{y(u)},
\end{align*}
where $\nu_{\tau}(u) = \int_{u}^{\tau}S(t)dt$. For the cumulative incidence of a type-$j$ event, the influence function is:
\begin{align*}
    \varphi_{i}(t) = - F_{j}(t) \int_{0}^{t}\frac{dM_{i}(u)}{y(u)} + \int_{0}^{t}\frac{F_{j}(u)dM_{i}(u)}{y(u)} + \int_{0}^{t}\frac{S(u)dM_{i,j}(u)}{y(u)}. 
\end{align*}
Here $F_{j}(t)$ is the cumulative incidence of type-$j$ events, and $M_{i,j}$ is a counting process martingale specific to type-$j$ events:
\begin{align}
    M_{i,j}(t) = N_{i,j}(t) - \int_{0}^{t}Y_{i}(u)dA_{j}(u),
\end{align}
where $N_{i,j} = \Ibb(U_{i} \leq t, \delta_{i} = j)$, and $A_{j}(u)$ is the cumulative hazard of type-$j$ events. 


\section{Mediation for Time-to-Event Outcomes using Pseudo-Values}
\label{sec:meth}

In this section, we join the elements reviewed in section 2 in order to describe the pseudo-values approach to mediation analysis for time-to-event outcomes. By leveraging pseudo-values, we are able to postulate linear models for the time-to-event estimands. With use of the models in (\ref{eqn:outcome-model}) and (\ref{eqn:mediator-model}), we can decompose the total effect in the natural direct and indirect (mediated) effects. We also discuss software implementations. 

\subsection{Step-by-Step Procedure}

The following steps carry out mediation for time-to-event estimands using pseudo-values:

\begin{flushleft}
\textbf{Step 1: Calculate Full Sample Estimate}
\par\end{flushleft}

\begin{quote}
    Calculate $\hat{\theta}$ the estimate for the entire sample (i.e. not separately in each treatment arm) at a chosen time $\tau$. Use the Kaplan-Meier estimator to obtain the survival probability estimate $\hat{S}(\tau)$ and RMST estimate $\hat{\mu}(\tau)$. The Aalen-Johansen estimator can be used to estimate the cumulative incidence function $\widehat{F}_d(\tau)$ for a type-$d$ event.
\end{quote}

\begin{flushleft}
\textbf{Step 2: Generate Pseudo-Values}
\par\end{flushleft}
\begin{quote}
Generate the pseudo-values $Y_i$ for each observation $i$ using the overall estimate $\hat{\theta}$ from step 1. This can be done either by using the leave-one-out estimator:
\begin{align}
    Y_{i} = n\hat{\theta} - (n-1)\hat{\theta}_{(-i)},
\end{align}
or with the influence function approximation to the pseudo-value from (\ref{eqn:pseudo-approx}). 
\end{quote}

\begin{flushleft}
\textbf{Step 3: Fit Linear Regression Mediation Models}
\par\end{flushleft}
\begin{quote}
First fit a regression model for the mediator:
\begin{equation}
    \hat{\Ebb}[M|A] = \hat{\alpha_0} + \hat{\alpha}_AA. 
\end{equation}
If treatment $A$ is randomized, it is not necessary to include confounding variables in this model. Otherwise, all confounders of the treatment-mediator relationship need to be included in the mediator model.

Next, using the pseudo values $Y_i$ calculated in step 2, fit the linear (pseudo-value) outcome model:
\begin{equation}
    \hat{\Ebb}[Y|A,M] = \hat{\beta}_0 + \hat{\beta}_AA + \hat{\beta}_MM + \hat{\beta}_CC  
\end{equation}

Any confounders ($C$) of the mediator-outcome relationship must be included in the pseudo-value outcome model above in order to identify the causal mediation effects. The confounders of this relationship are not handled by randomization of treatment $A$ and should be adjusted for in almost all cases. In non-randomized settings, confounders of the treatment-outcome effect must also be included in this model.
\end{quote}

\begin{flushleft}
\textbf{Step 4: Combine Estimates}
\par\end{flushleft}
\begin{quote}
The coefficients from these models can then be combined in the same way as outlined in section 2.1 to obtain the causal mediation effects:
\begin{align}
    \text{Natural direct effect}:\quad \widehat{\text{NDE}} &= \hat{\beta}_A, \\
    \text{Natural indirect effect}:\quad \widehat{\text{NIE}} &= \hat{\alpha}_A\hat{\beta}_M, \\
    \text{Total effect}:\quad \widehat{\text{TE}} &=\hat{\beta}_A + (\hat{\alpha}_A\hat{\beta}_M).
\end{align}
One potential robustness check would be to estimate the total effect independently of these models (e.g. by taking the difference in survival probabilities or RMSTs between either treatment group in a randomized setting). Since this estimates the same quantity as $\widehat{TE}$ above, it should be similar to the approach above if the models are not grossly misspecified.

\end{quote}

\begin{flushleft}
\textbf{Step 5: Inference}
\par\end{flushleft}
\begin{quote}
Inference can be conducted by bootstrapping steps 2-4 to obtain confidence intervals and p-values. For each bootstrap sample, it is important to re-calculate the pseudo-values (step 2) in addition to the mediation model fits. Calculating the pseudo-values for each bootstrap sample is computationally intensive and the influence function approximation of the pseudo-values can reduce the computational time substantially when bootstrapping. 

Alternatively, the closed form expression for the exact standard error of the indirect effect based on a second-order Taylor expansion \cite{aroian1947probability,mackinnon2002comparison} can be applied:
\begin{equation}
    \hat{\sigma}_{\text{NIE}} = \sqrt{\hat{\alpha}_A^2\,\hat{\Vbb}(\hat{\beta}_M) + \hat{\beta}_M^2\,\hat{\Vbb}(\hat{\alpha}_A) + \hat{\Vbb}(\hat{\beta}_M)\hat{\Vbb}(\hat{\alpha}_A)}.
    \label{eqn:exactCI}
\end{equation}
Yet another popular option is to use the multivariate delta method  (first-order Taylor expansion) to estimate the sampling distribution of the product of the regression coefficients in order to derive approximate p-values and confidence intervals \cite{cheng2021estimating, vanderweele2009conceptual}. Bayesian inference for mediation is also well formulated \cite{yuan2009bayesian}. 
\end{quote}

The above approach considers a continuous mediator $M$, but as discussed in the next section 3.2, formal mediation software can be used to handle different types of mediators (e.g. binary, nominal, count, etc.) with a pseudo-value outcome model.

\subsection{Software}

Due to the simplification of the problem via pseudo-values, standard statistical software can be used to obtain estimates for these mediation analyses. We provide herein some guidance and recommendations based on our experience. 

One can always obtain the pseudo-values for each observation by contrasting the leave-one-out (jackknife) estimate with the overall estimate (i.e. $Y_i=n\hat{\theta}-(n-1)\hat{\theta}_{(-i)}$). However, as the sample size increases, calculation of the pseudo-values via jackknifing becomes computationally burdensome. Pseudo-values targeting the survival probability or the RMST can be calculated via the \texttt{pseudo()} function from the \texttt{R} package \texttt{survival} \cite{survival-package}. The function \texttt{pseudo()} leverages the influence function approximation discussed in section 2.4 to accelerate the computation. In samples of even moderate size (e.g.\@ $N = 50$), the influence function-based approximation to the pseudo-value differs only negligibly from the jackknife version (see section 4.3.3). Example code for calculating pseudo-values for both the survival probability and RMST at $\tau=2$ is:
\begin{alltt}
    library(survival)
    km_fit <- survfit(Surv(time,event) ~ 1, data = df) # Kaplan-Meier Estimator
    pseudo(fit = km_fit,times = 2,type = "surv")
    pseudo(fit = km_fit,times = 2,type = "rmst")
\end{alltt}
As the \texttt{survival} package does not yet appear to support pseudo-value calculation for the cumulative incidence, we implemented the calculation in the \texttt{SurvUtils} package \cite{survutils-package}. Example code for calculating cumulative incidence pseudo-values is:
\begin{alltt}
    library(SurvUtils)
    GenPseudo(data = data, type = "cic", tau = 2)
\end{alltt}
Here \texttt{data} is a data frame containing columns \texttt{time} and \texttt{status}, where \texttt{status} is coded as 0 for censoring, 1 for the event of interest, and 2 for the competing risk. \texttt{GenPseudo} can also calculate pseudo-values for the survival probability (\texttt{type = "prob"}) and RMST (\texttt{type = "rmst"}). 

Once the pseudo-values are constructed, standard linear regression software (e.g. \texttt{lm()} in \textsf{R}) can be used to fit the models for the pseudo-value outcome and a model for the mediator as in step 3 above. Then the outputs from these models can be used to combine coefficients as in step 4 and 95$\%$ CI could then be constructed via the bootstrap described above or using the closed form expression for the variance (\ref{eqn:exactCI}).

Another option is to input the pseudo-values into existing mediation software (i.e. CMAverse), making sure to specify the pseudo-value outcome as continuous rather than time-to-event. For example, once the pseudo values are calculated and appended to the data as \texttt{df\$pseudo}, one could apply the \texttt{cmest} function from the CMAverse package:

\begin{alltt}
    library(CMAverse)
    cma_mod <- cmest(data = df, model = "rb", EMint = F, 
                     outcome = "pseudo", exposure = "A", mediator = "M",
                     mreg = list("linear"), yreg = "linear", mval = list(0),
                     estimation = "paramfunc", inference = "delta")
\end{alltt}

Then the NDE and NIE with their corresponding 95\% CIs and p-values can be extracted from the \texttt{cma\_mod}
object via \texttt{summary(cma\_mod)}. Using existing mediation software such as \texttt{CMAverse} has the advantage of being able to handle more complicated scenarios such as non-continuous mediators, treatment-mediator interactions, multiple mediators, confounders affected by treatment, etc. Furthermore, these packages guarantee appropriate statistical inference. However, specifying \texttt{inference="bootstrap"} in CMAverse is not appropriate for the pseudo-value method as the computation of the pseudo-values needs to done for each bootstrapped sample. For more on the CMAverse, see \cite{shi2021cmaverse}. The CMAverse is just one of many available packages that employ mediation analyses \cite{steen2017medflex,bach2022doubleml}. 

A data application on a simulated dataset is given in the appendix. A more thorough \texttt{R} code vignette implementing the approach from start to finish for various estimands and time points on a simulated dataset for readers to adopt to their own data applications is given in the supplementary material.   

\section{Simulation Studies}

The goal of these simulation studies is to demonstrate that the pseudo-value approach outlined in section 3 provides unbiased estimates of the NDE and NIE for the various survival estimands defined in section 2.2. We also assess if the method obtains adequate inferences by assessing type-I error control and coverage. Lastly, we investigate computation time and accuracy of the influence function approximation to the pseudo-values. To foster reproducibility, the code to implement these simulations is included in the supplementary materials.

\subsection{Simulation Setup}

For each of $2N$ subjects ($N$ per arm, $N\in \{50,100,200$\}), let $A \in \{0, 1\}$ denote the treatment arm, and $M$ the measurement of a continuous mediator. The mediator is drawn from a normal distribution that is dependent on the value of $A$:
\begin{align*}
    M|(A=a) \sim N(\mu_{a}, 1),\quad a \in \{0, 1\}
\end{align*}
where the mean $\mu_{a}$ depends on the treatment arm. In our simulations, we let $\mu_{0} = 0$ and $\mu_{1} = -1$. The time-to-event (TTE) is simulated from an exponential distribution, parameterized as:
\begin{align}
    f(t;\lambda) = \lambda e^{-\lambda t}.
    \label{eqn:exp-dens}
\end{align}
The rate parameter depends on the treatment and the mediator: 
\begin{align}
    \lambda =  \exp\big(\beta_{0} + A\beta_{A} + M\beta_{M}\big).
    \label{eqn:lambda-model}
\end{align}
The simulation addresses the following 4 cases:
\begin{enumerate}
    \item No effect, $\beta_{A} = 0$ and $\beta_{M} = 0$. 
    \item Direct effect only, $\beta_{A} \neq 0$ and $\beta_{M} = 0$.
    \item Indirect effect only, $\beta_{A} = 0$ and $\beta_{M} \neq 0$.
    \item Both direct and indirect effects, $\beta_{A} \neq 0$ and $\beta_{M} \neq 0$. 
\end{enumerate}
When there are effects, we parameterize them as follows. For $\beta_{0}$, let:
    \begin{align*}
        \beta_{0} = \ln\left(\frac{1}{k}\right),
    \end{align*} 
    which results in an expected TTE of $k$ years in the no-effect setting. For $\beta_{A}$, let:
    \begin{align*}
        \beta_{A} = \ln\left(\frac{k}{k + 1}\right).
    \end{align*}
    This models a treatment that adds 1 year of TTE in the direct effect only setting. For $\beta_{M}$, let:
    \begin{align*}
        \beta_{M} = \ln\left(\frac{k + 1}{k}\right).
    \end{align*}
This models a treatment where a patient gains 1 year of TTE when the mediator is reduced by 1 unit, as by treatment, in the indirect only setting. For our simulations we consider $k=3$.

For censoring, we simulate independent $C \sim \text{Exp}(\lambda_{C})$ with
    \begin{align*}
        \lambda_{C} = \frac{\pi_{C}}{1 - \pi_{C}}\lambda_{0},
    \end{align*}
where $\pi_{C}$ is the target censoring fraction (e.g.\@ $\pi_{C} = 0.2$) and $\lambda_{0}$ is a common event rate for the simulation setting. For example, under the no effect case, $\lambda_{0} = \exp(\beta_{0}) = k^{-1}$. Under the direct effect only and indirect effect only case $\lambda_{0} = (k+1)^{-1}$. Under both direct and indirect effects, $\lambda_{0} = k(k+1)^{-2}$.  

Overall, each patient's simulated data is:
    \begin{align*}
        \mathcal{D}_{i} = (A_{i}, M_{i}, T_{i}, C_{i}, U_{i}, \delta_{i}),
    \end{align*}
    where $U_{i} = \min(T_{i}, C_{i})$ and $\delta_{i} = \Ibb(T_{i} \leq C_{i})$. The pseudo-values are calculated from $(U_{i}, \delta_{i})$. 

In some scenarios, we evaluate the impact of competing risks that hinder the ability to observe the event and bias the Kaplan-Meier estimate of the cumulative incidence function $F_T(t)$. In these scenarios, we now distinguish the time to the event of interest $T$ from the time to the competing risk $D$ (e.g. death). $T$ and $D$ follow independent exponential distributions:
\begin{align*}
    T \sim \text{Exp}(\lambda_{T}) \perp D \sim \text{Exp}(\lambda_{D}),
\end{align*}
with respective rates $\lambda_{T}$ and $\lambda_{D}$. The cumulative incidence of the event of interest $T$ at time $\tau$ is:
\begin{align*}
    F_{T}(\tau) \,= \int_0^\tau \lambda_T\,\Pbb\!\left\{\textrm{min}(T,D)\ge t\right\}\!dt \,=\, \frac{\lambda_T}{\lambda_T+\lambda_D} \left(1-e^{-(\lambda_T+\lambda_D)\tau} \right)\!. 
\end{align*}
The rate of the outcome event of interest will depend on the treatment and biomarker as above:
\begin{align*}
    \lambda_{T} = \exp(\beta_{0} + A\beta_{A} + M\beta_{M}). 
\end{align*}
While we could specify a similar model for $D$, for simplicity, we assume $\lambda_{D}$ is a constant. 
%
The overall data for the competing risks analyses are:
\begin{align*}
    \mathcal{D}_{i} = (A_{i}, M_{i}, T_{i}, D_{i}, C_{i}, U_{i}, \delta_{i}), 
\end{align*}
where $C_{i}$ is the censoring time, $U_{i} = \min(C_{i}, T_{i}, D_{i})$ and:
\begin{align*}
    \delta_{i} = \begin{cases}
        0, & \text{ if } U_{i} = C_{i}, \\
        1, & \text{ if } U_{i} = T_{i}, \\
        2, & \text{ if } U_{i} = D_{i}.
    \end{cases}
\end{align*}

\subsection{Mediation Estimands}


Given this setup, our target estimands are the total effect (TE), natural direct effect (NDE), and natural indirect effect (NIE) for the survival probability, RMST, and cumulative incidence function taking into account the competing risk. 

For an exponentially distributed random variable $T$, the true survival probability at time $\tau$ is:
\begin{align*}
    S(\tau) = e^{-\lambda\tau}.
\end{align*}
Therefore, the TE, NDE, and NIE on the survival probability scale at time $\tau$ are:
\begin{align*}
    \text{TE}(\tau) = \Ebb_{M_1}\big\{S(\tau; A=1, M_{1})\big\} - \Ebb_{M_0}\big\{S(\tau; A=0, M_{0})\big\},\\
     \text{NDE}(\tau) = \Ebb_{M_0}\big\{S(\tau; A=1, M_{0})\big\} - \Ebb_{M_0}\big\{S(\tau; A=0, M_{0})\big\},\\
     \text{NIE}(\tau) = \Ebb_{M_1}\big\{S(\tau; A=1, M_{1})\big\} - \Ebb_{M_0}\big\{S(\tau; A=1, M_{0})\big\}.
\end{align*}
Note that these integrals are with respect to the mediator $M$, since based on (\ref{eqn:lambda-model}), different values of $M$ lead to different survival curves.

Given the exponentially distributed random variable $T$, the true RMST at a given time $\tau$ is:
\begin{align*}
    \mu(\tau; \lambda) = \int_{0}^{\tau} e^{-\lambda t}dt = \frac{1}{\lambda}\big(1 - e^{-\lambda \tau}\big),
\end{align*}

which we can use to define our TE, NDE, and NIE as follows:
\begin{align*}
     \text{TE}(\tau) = \Ebb_{M_1}\big\{\mu(\tau; A=1, M_{1})\big\} - \Ebb_{M_0}\big\{\mu(\tau; A=0, M_{0})\big\},\\
     \text{NDE}(\tau) = \Ebb_{M_0}\big\{\mu(\tau; A=1, M_{0})\big\} - \Ebb_{M_0}\big\{\mu(\tau; A=0, M_{0})\big\},\\
     \text{NIE}(\tau) = \Ebb_{M_1}\big\{\mu(\tau; A=1, M_{1})\big\} - \Ebb_{M_0}\big\{\mu(\tau; A=1, M_{0})\big\},
\end{align*}
For the cumulative incidence function in the presence of competing risk $D$, the mediation estimands are defined as follows:
\begin{align*}
    \text{TE}(\tau) = \Ebb_{M_1}\big\{F_{T}(\tau; A=1, M_{1})\big\} - \Ebb_{M_0}\big\{F_{T}(\tau; A=0, M_{0})\big\},\\
    \text{NDE}(\tau) = \Ebb_{M_0}\big\{F_{T}(\tau; A=1, M_{0})\big\} - \Ebb_{M_0}\big\{F_{T}(\tau; A=0, M_{0})\big\},\\
    \text{NIE}(\tau) = \Ebb_{M_1}\big\{F_{T}(\tau; A=1, M_{1})\big\} - \Ebb_{M_0}\big\{F_{T}(\tau; A=1, M_{0})\big\}.
\end{align*}

For all the above, $M_{0} \sim N(0, 1)$ and $M_{1} \sim N(-1, 1)$. Since all these estimands are expressed in terms of expectations over a normal random variable (i.e.\@ the mediator conditional on the value of $A$), we can use the Gauss-Hermite quadrature \cite{heath2002} to numerically calculate the true values. Using these true values, we assess bias and coverage for this truth where inference is conducted using the delta method approximation for the NIE to obtain 95\% CI and p-values.

\subsection{Results}

The preceding subsections described our generative model and the formulas for calculating the true total, natural direct, and natural indirect effects (i.e.\@ the mediation estimands). This section describes the operating characteristics of inference on those estimands using pseudo-value-based inference. We consider various sample sizes, $N \in \{50, 100, 200\}$ per-arm, evaluation time points, $\tau \in \{2, 3, 4\}$, and generative models, as specified by whether $(\beta_{A}, \beta_{M})$ are non-zero. To avoid redundancy, we sometimes present only a subset of results when the operating characteristics were similar across multiple scenarios. An extensive table summarizing the operating characteristics of pseudo-value-based inference across all 324 estimands/cases considered is provided in the supplementary materials.

\subsubsection{Bias}

\begin{figure}[H]
\centering
{\includegraphics[width=1\columnwidth]{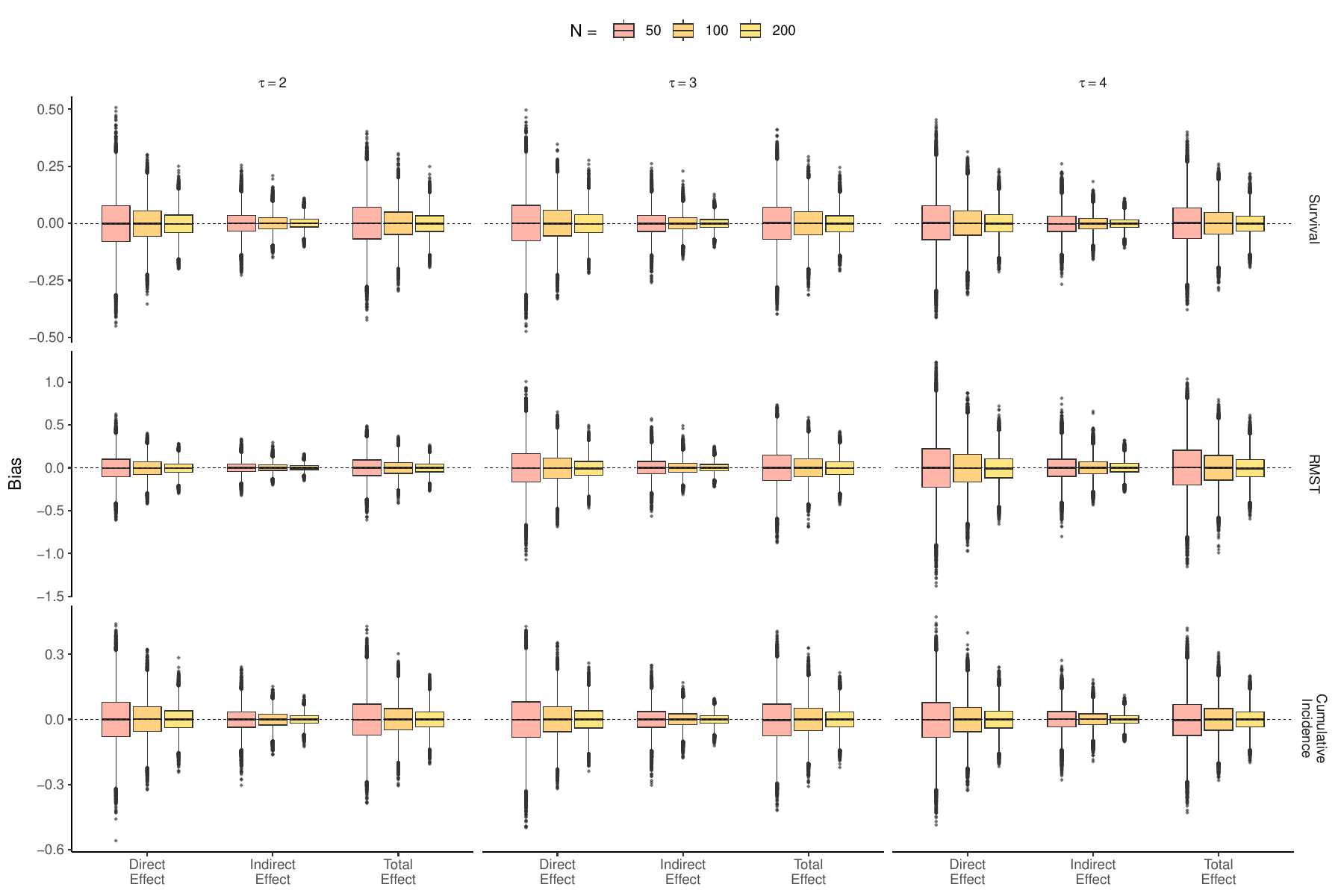}}
\caption{\textbf{Across estimands, time points, and sample sizes, pseudo-value based inference provides unbiased estimates of mediation estimands.} Bias was calculated relative to the true values of the estimands across $\tau \in \{2, 3, 4\}$ and sample sizes per-arm of $N \in \{50, 100, 200\}$ in the presence of both direct and indirect effects ($\beta_A\neq0$ and $\beta_M\neq0$).}
\label{fig:bias}
\end{figure}

The pseudo-value approach was unbiased for the true NIE, NDE, and TE across all estimands, time points $\tau$, sample sizes $N$, and whether or not the direct or indirect effects were present. Figure \ref{fig:bias} presents the bias for the three mediation estimands on the survival probability, RMST, and cumulative incidence curve scales, in the case where the indirect and direct effect are both present ($\beta_A\neq0$ and $\beta_M\neq0$ in equation \ref{eqn:lambda-model}). The bias was evaluated across 10,000 simulation replicates. As expected, the variance of the effect size distribution decreases as the sample size increases. Overall, these results confirm the validity of the general approach outlined in section \ref{sec:meth} for estimating mediation effects via pseudo-values. As demonstrated in the supplementary material, estimation was likewise unbiased in the absence of any effect, and when only one of the direct or indirect effects was present.


\subsubsection{Type I Error \& Coverage}

\begin{figure}[H]
\centering
{\includegraphics[width=0.9\columnwidth]{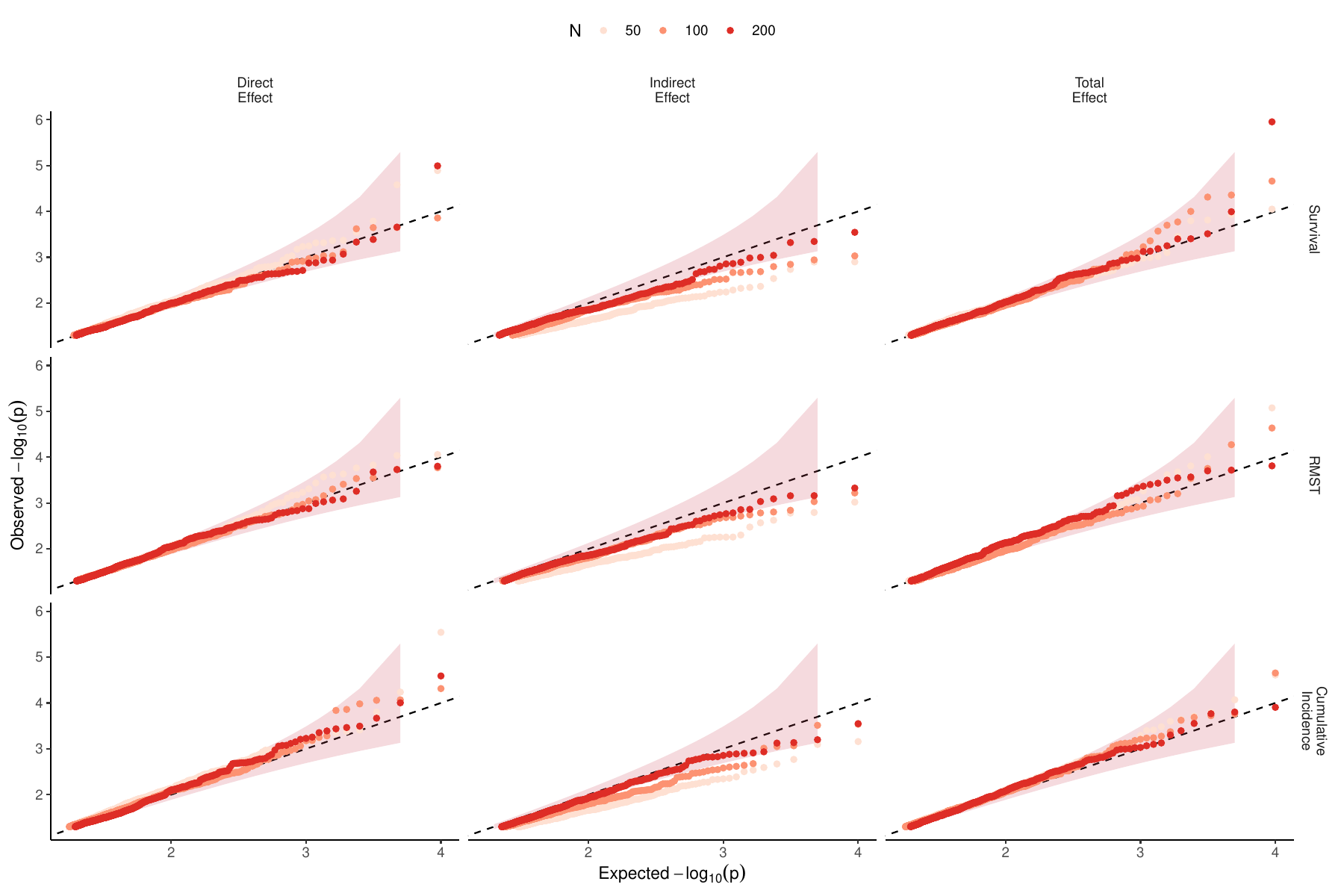}}
\caption{\textbf{P-values for pseudo-value based mediation effects are uniformly distributed under the null.} Shown are uniform quantile-quantile plots for the observed vs.\@ expected p-value distributions under the null scenario ($\beta_A=0$ and $\beta_M=0$). Performance is evaluated for the survival probability, RMST, and cumulative incidence curve mediation estimands at $\tau=2$ and sample sizes per-arm of $N \in \{50, 100, 200\}$. Adherence to the diagonal indicates that the p-values are uniformly distributed under $H_{0}$. The shaded area is the 95\% confidence band. The presence of p-values systematically below the confidence band indicates conservative inference, while the presence of p-values systematically above the confidence band would indicate type I error inflation.}
\label{fig:t1e}
\end{figure}

We next validated that inference on the total, natural direct, and natural indirect effects properly controls the type I error under the null scenario of $\beta_A=0$ and $\beta_M=0$. As demonstrated by the uniform quantile-quantile (QQ) in Figure \ref{fig:t1e}, p-values for the total effect and natural direct effect were uniformly distributed under $H_{0}$ at all sample sizes. For the natural indirect effect, inference was slightly conservative at smaller sample sizes, as indicated by deflection of the curve below the diagonal. While conservative inference is not a problem for validity, as the type I error is maintained below its nominal level, this could have implications for power. Note, however, that this conservatism fades as the sample size increases, and has essentially vanished by the time $N = 200$. At the standard type I error level of $\alpha = 0.05$, the empirical type I error was 0.05 for the total effect, 0.05 for the natural direct effect, and 0.04 for the natural indirect effect averaged across all cases considered. We also checked the empirical coverage of 95\% confidence intervals (CIs). As shown in the supplemental materials, coverage for the total and natural direct effects was close to nominal for all sample sizes. Coverage for the natural indirect effect was slightly elevated at small samples sizes, but approached nominal as the sample size increased. 

\subsubsection{Influence Function Approximation}

\begin{figure}[H]
    \centering

    \begin{subfigure}{0.95\textwidth}
        \centering
        \caption{\textbf{Pseudo-value Accuracy}}
        \includegraphics[width=\textwidth]{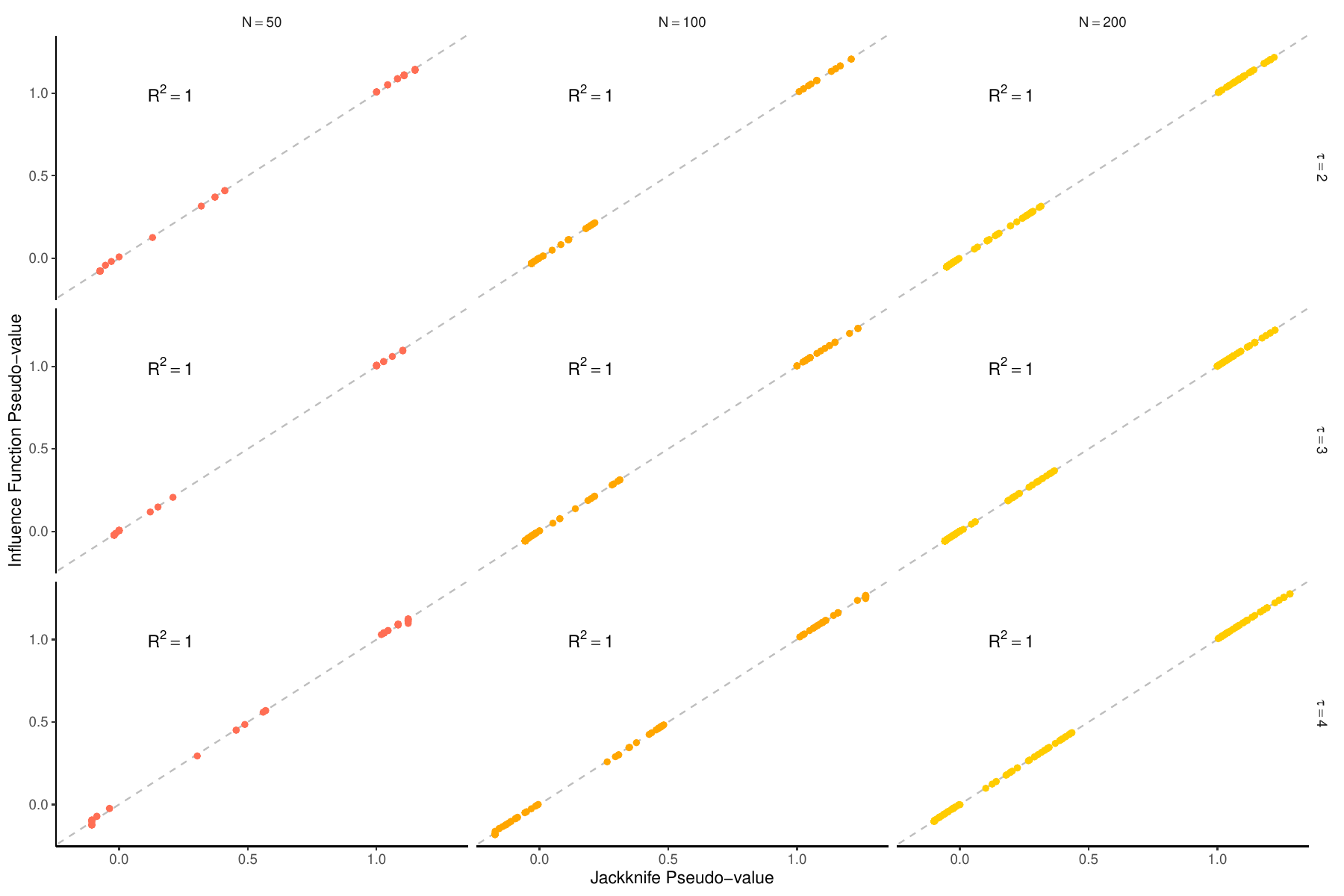}
    \end{subfigure}

        \begin{subfigure}{0.95\textwidth}
        \centering
        \caption{\textbf{Runtime Scaling}}
        \includegraphics[width=\textwidth]{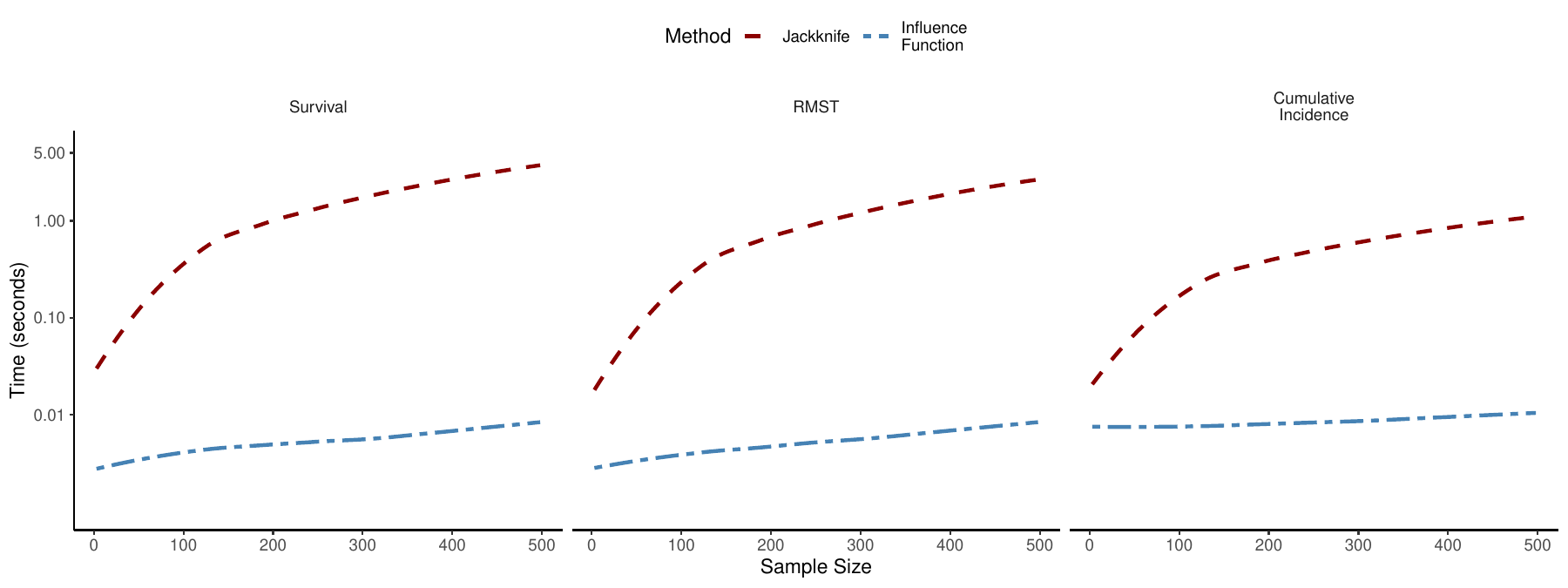}
    \end{subfigure}
    
    \caption{\textbf{Influence function approximation to the jackknife accelerates pseudo-value calculation without compromising accuracy.} (a) Comparison of influence function versus jackknife calculation of the pseudo-value for the cumulative incidence function across sample sizes and time points. Influence function-based pseudo-values were calculated using \texttt{SurvUtils} \cite{survutils-package}. (b) Runtime by sample size for pseudo-value calculation by means of jackknifing vs.\@ the influence function approximation as a function of sample size $N$.}
    \label{fig:run-time}
\end{figure}

Calculating pseudo-values by means of jackknifing becomes increasingly expensive as the sample size increases. This is because the target of inference (e.g.\@ cumulative incidence function) must be calculated first based on the full sample, then omitting each subject in turn. The computational cost can become prohibitive when, for example, pseudo-value calculation is being used in conjunction with bootstrapping. As discussed in section \ref{sec:influence-functions}, when the influence function of an estimand is known, an approximate means of calculating the pseudo-value is available, which avoids the need for jackknifing. We validated empirically that the influence function-based approximation is exceptionally accurate. Figure \ref{fig:run-time}(a) illustrates, for the case of the cumulative incidence function, that the $R^2$ between the influence function and jackknife pseudo-values was, to 2 significant digits, 1.00 across all time points and sample sizes evaluated. Similar results for the survival probability and RMST are presented in \textbf{Supplemental Figure 1}. Results for the survival probability and RMST were nearly identical. Figure \ref{fig:run-time}(b) validates that, as expected, the influence function-based calculation is substantially faster, and that the runtime advantage of the influence function versus jackknife calculation grows as the sample size increases.

\section{Example from a Clinical Trial for Multiple Sclerosis}

We explore our proposed method in a clinical trial for Multiple Sclerosis (MS) where we investigate how the treatment effect on time-to-relapse is mediated by T1-gadolinium-enhancing (T1-gd) lesions from Magnetic Resonance Imaging (MRI). T1-gd lesions are objective evidence of the presence of acute inflammation in the brain; relapses are the symptoms resulting from this inflammation. T1-gd enhancing lesions have been known to be strongly correlated with relapses and are therefore often used as surrogate endpoints in phase II proof of concept and dose-finding studies which inform the potential treatment effect on relapses in the phase III setting \cite{sormani2013mri}. However, formal causal analyses of this relationship are lacking in the MS literature. This mediation analysis aims to add supportive, causal evidence that T1-gd lesions are on the causal pathway between treatment and a clinical relapse.

The clinical trial considered was the PARADIGMS study (NCT01892722) \cite{chitnis2018trial}, which demonstrated that fingolimod (FTY) significantly reduced relapses versus interferon $\beta$-1a (INF-$\beta$1a) in pediatric MS patients over the roughly 2 years of follow-up. For the mediation analysis, our outcome is time-to-relapse in the second year of MS treatment and our mediating variable is the log number of T1-gd lesions at year 1. Because the mediator is collected after the first year of treatment, we cannot use time from treatment initiation as the time 0 in our time-to-event analysis. Instead, to keep events in sequence, we re-baseline time-to-relapse starting at year 1, the time-point for which we collected the biomarker (i.e.~the number of T1-gd lesions at year 1). Since relapses are a recurrent event, all patients can contribute to this analysis, i.e. no censoring is needed for those who had a relapse already in the first year because all patients can experience a relapse in the second year of treatment. In our pseudo-value outcome model, we adjusted for age, number of relapses in the last year pre-randomization, normalized brain volume, sex, duration since first symptoms, and previous treatment usage all measured at baseline (i.e. at or slightly before the first dose of randomized treatment). The mediator model for log T1-gd lesions did not adjust for any confounders since treatment is randomized.





The estimates of the natural direct, indirect, and total effect are presented in Table 1 on both the survival probability and RMST scale. The analysis detects a mediated pathway, thereby revealing that the treatment effect of fingolimod versus interferon on relapses in the second year can be, in part, explained by the treatment’s effect on reducing the number of T1-gd lesions in the first year. Approximately 25$\%$ of the treatment’s effect on relapses in the second year is found to be mediated by the reduction in T1 lesion numbers at year 1. It is of note that, although not identical, the proportion mediated is estimated to be similar on both the survival probability or the RMST scale. Figure 1 depicts the cumulative incidence curve (i.e. 1 - Kaplan-Meier curve) estimates for time-to-relapse in the 2nd year along with the total effect at $\tau=2$ decomposed into its direct and mediated pathways.

\begin{table}[H]
\centering
\begin{tblr}{
  column{even} = {c},
  column{3} = {c},
  column{5} = {c},
  cell{1}{2} = {c=2}{},
  cell{1}{4} = {c=2}{},
  hline{1-3,6-7} = {-}{},
}
                        & Survival Probability &         & RMST (years)      &         \\
Estimand                & Estimate (95\% CI)   & p-value & Estimate (95\% CI) & p-value \\
Natural Indirect Effect & 0.07 (0.02, 0.14)    & 0.022    & 0.05 (0.02, 0.10)  & 0.023   \\
Natural Direct Effect   & 0.23 (0.03, 0.42)    & 0.009    & 0.15 (0.05, 0.23)  & <0.001   \\
Total Effect            & 0.30 (0.11, 0.48)    & <0.001   & 0.20 (0.10, 0.29)  & <0.001   \\
Proportion Mediated     & 24\% (6\%, 76\%)     &     & 26\% (9\%, 57\%)  &    
\end{tblr}
\caption{Mediation Analysis for the proportion of the treatment effect (FTY vs. INF-$\beta$1a) on time-to-relapse after year 1 mediated by T1-gd enhancing lesions at year 1 for both the survival probability and restricted mean survival time.}
\end{table}

Estimates were calculated with the the influence function approximation to the pseudo-value for both the survival probability and RMST using the \texttt{pseudo()} function. The mediation pathways were estimated by combining the estimates using two \texttt{lm()} fits in \texttt{R}, one with the mediator as the outcome regressed on treatment and one with the pseudo-value as the outcome regressed on treatment and covariates detailed above. Inference was conducted by bootstrapping this procedure with 1,000 replicates - although delta method CIs provided similar results. Code is provided in the supplementary material.

\begin{figure}[H]
\centering
{\includegraphics[width=1\columnwidth]{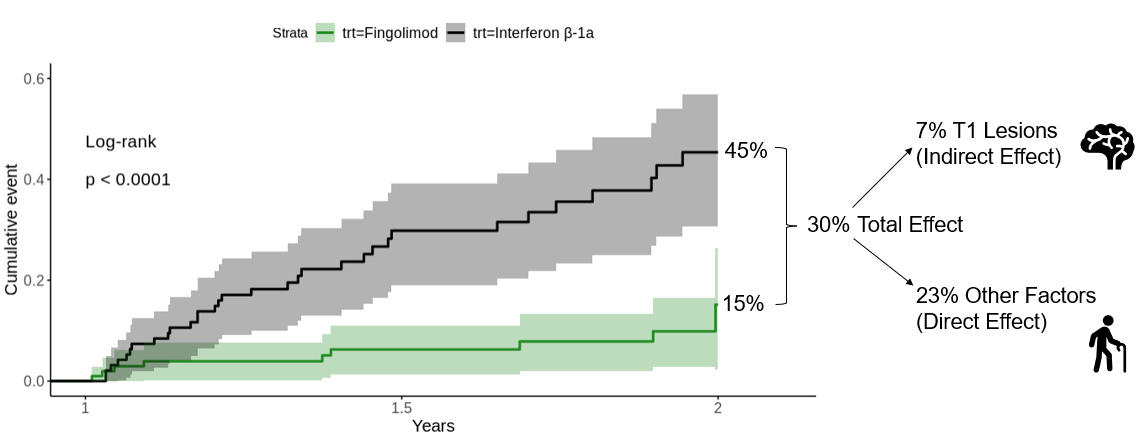}}
\caption{Time-to-relapse cumulative incidence curves under FTY and INF-$\beta$1a with survival probabilities decomposed into the direct effect and T1 lesion mediated indirect effect.}
\label{fig:mediation-graph}
\end{figure}

From a biological point of view, if we had continuous measurements of T1-gd lesions we would expect the proportion mediated to be higher. For practical reasons, however, we only had an MRI scan at year 1. Despite this limitation, the analysis detected that a significant proportion of the effect on relapses in the second year of the study is mediated by the effect on MRI lesions at year 1. This is supportive evidence of the causal relationship between T1-gd lesions and relapses and a testimony of the importance of fingolimod's superior ability to reduce T1-gd lesions (compared to IFN-$\beta$1a) for the suppression of relapses.  Furthermore, our analysis confirms the causal relationship between T1-gd lesions and relapses, which can be be used as an additional justification for using T1-gd lesions as a surrogate outcome for relapses in future MS studies.

\section{Discussion}

In this manuscript, we developed mediation analysis for survival endpoints by means of pseudo-values. We illustrated that pseudo-values are straightforward to calculate, for example in \texttt{R} using the \texttt{survival} or \texttt{SurvUtils} packages, and once obtained, can be analyzed using familiar linear models, as implemented by \texttt{lm()} or in the \texttt{CMAverse} package. Through extensive simulation studies, we verified that estimation is unbiased and inference is valid for multiple estimands, including the natural direct, natural indirect, and total effects, as quantified in terms of the survival probability, the restricted mean survival time, and the cumulative incidence in the presence of competing risks. We further confirmed that influence function approximations to the pseudo-values are accurate and computationally efficient. We applied our approach to investigate the effect of fingolimod on time-to-relapse, as mediated by the number of T1-gd lesions detected at year 1, finding that there was significant evidence of mediation. 

Mediation for survival outcomes has long been less-developed than for other types of outcomes. We believe this is at least in part because the applied analyst’s standard tool for regression analysis of survival outcomes, the Cox model, is poorly suited to the task of mediation analysis. In mediation analysis, the general goal is to decompose a total effect into direct and indirect (or mediated) components. However, the hazard ratio estimated by the Cox model is not collapsible. Consequently, the difference between the total and direct effect estimates from a Cox model conflates mediation with non-collapsibility. Moreover, because proportional hazards cannot simultaneously hold for the Cox models with and without the mediator (if a mediated effect is present), at least one of the total effect or the direct effect estimates will generally be biased. The reason for even contemplating Cox models when performing mediation analysis with time-to-event outcomes is not due to their suitability for mediation analysis, but rather because they can accommodate censoring.

In this paper, we have developed an approach to mediation analysis, based on pseudo-values, which effectively decouples the censoring problem from estimation of causal effects. The strength of this approach is that after having computed the pseudo-values, essentially all existing methods for performing mediation analysis on continuous outcomes can be applied. Thus, linear regression, which provides collapsible effect size estimates and imposes no proportional hazards assumptions, may be used to obtain unbiased effect estimates, as we have demonstrated. The primary limitation of our approach is that pseudo-values are only applicable to estimating quantities that can be expressed as an expectation of the survival time. Thus, pseudo-values are not directly applicable to estimating an effect measure such as the hazard ratio, although this is arguably not a drawback insofar as the hazard ratio is difficult to interpret for the reasons discussed above. Additionally, regression analysis of pseudo-values remains relatively unknown despite having been introduced in 2003 \cite{andersen2003}. It is therefore expected that a substantial explanation of pseudo-values as an existing, theoretically grounded tool for covariate adjustment in survival analysis may be required in applied work.

The PARADIGMS trial analysis demonstrates the utility of the approach with evidence that fingolimod's effect on delaying time-to-relapse is partly due to reducing T1-gd lesions, aligning with the current understanding of multiple sclerosis pathobiology. The analysis serves as a causal compliment to existing research on the subject. Statistically, the total effect decomposition (TE = NIE + NDE) was additive both on the survival probability (0.30 = 0.07 + 0.23) and RMST (0.20 = 0.05 + 0.15) scales. Existing mediation methods for time-to-event data that decompose HRs lack this clear additive decomposition. Additive decompositions are more intuitive than multiplicative ones; for instance, it is straightforward to infer the proportion mediated. Our approach accomplishes an additive decomposition because, as a starting point, our total effects are differences in summary measures and not ratios.  Also, the proportion mediated was similar for both the survival probability (24\%) and RMST (26\%). This suggests the mediation analysis is robust to the summary measure chosen. More generally, clinical trials serve as fertile ground for mediation analyses as randomization guarantees no unobserved confounding of the exposure-outcome and exposure-mediator relationships. Of course, the mediator-outcome relationship cannot be protected through randomization, and therefore the consideration, collection, and control of confounders for this relationship is essential, even in randomized clinical trials. 

An important aspect of mediation analysis with survival outcomes is that the exposure, mediator, and outcome must be ordered consecutively in time. Thus, in our analysis of the PARADIGMS trial, we considered the effect of a biomarker measured at 1 year after treatment initiation on subsequent survival. While this is sensible from the standpoint that the treatment needs time to affect the mediator, it complicates the interpretation of the causal effect by essentially conditioning on survival until the biomarker could be measured. Luckily, we were able to maintain the full sample size in this analysis as having a relapse in the first year does not preclude a patient from further observation (because MS relapses are recurrent events). However, this would not be possible if the event was death, for example. This discussion highlights that more work is needed to develop interpretable methods, like the one described herein, for estimating causal effects on time-to-event outcomes where measurement of the mediator may itself be subject to censoring. 

\printbibliography

\begin{flushleft}
\textbf{Acknowledgments}
\par\end{flushleft}
Many thanks to the members of the Novartis causal mediation focus area sub-team for helpful comments as the paper developed. We also appreciated feedback during the Neuroscience statistics knowledge sharing meeting and causal inference in practice seminar at Novartis. Special thanks to Ariel Chernofsky for guidance on implementation of the procedure. Lastly, a huge thank you to the patients and physicians who participated in the PARADIGMS clinical trial.

\newpage

\begin{flushleft}
\textbf{Appendix: R Code Example}
\end{flushleft}
\begin{alltt}
simulate_data <- function(
    N = 100, # number of subjects per arm
    k = 2, # expected TTT (years) in the no effect setting
    A_flag = TRUE, # logical value indicating whether to include direct effect
    M_flag = TRUE, # logical value indicating whether to include indirect effect
    mu_1 = -1, # expectation of M|(A=1)
    pi_c = 0.2 # proportion censored
)\{
  simdf <- data.frame(
    id = 1:(2*N), # patient id's
    A = rep(c(0, 1), N), # treatment assignment
    M = NA, # Mediator values
    time = NA, # TTE values
    C = NA, # Censoring time
    U = NA, # min(time, C)
    event = NA # event indicator
  )
  simdf$M <- rnorm(2*N, mu_1*simdf$A, 1) # Generate mediator values
  lambda <- 1/k * (k/(k+1))^(A_flag * simdf$A) * ((k+1)/k)^(M_flag * simdf$M)
  simdf$time <- rexp(2 * N, lambda) # Generate TTE
  lambda_0 <- 1/k * (k/(k+1))^A_flag * ((k+1)/k)^(-M_flag)
  lambda_c <- pi_c * lambda_0 / (1-pi_c) 
  simdf$C <- rexp(2*N, lambda_c) # Generate censoring
  simdf$U <- apply(simdf[c("time", "C")], 1, min)
  simdf$event <- 1 * (simdf$time <= simdf$C) # Generate observations
  return(simdf)
\}

# Simulate Data
set.seed(907)
df <- simulate_data(N = 200, k = 2)

# Pseudo-value mediation
library(survival)
km_fit <- survfit(Surv(time,event) ~ 1, data = df) # Kaplan-Meier Estimator
df$pseudo <- pseudo(fit = km_fit,times = 2,type = "surv")
# df$pseudo <- pseudo(fit = km_fit,times = 2,type = "rmst") # uncomment for rmst

library(CMAverse)
cma_mod <- cmest(data = df, model = "rb", EMint = F,
                 outcome = "pseudo", exposure = "A", mediator = "M",
                 mreg = list("linear"), yreg = "linear", mval = list(0),
                 estimation = "paramfunc", inference = "delta")
summary(cma_mod)
\end{alltt}

\end{document}